\documentclass[a4paper,11pt]{article}
\usepackage{jheppub} 
                     
\usepackage{hyperref}

\usepackage[utf8]{inputenc}
\usepackage{amsmath}
\usepackage{amsfonts}
\usepackage{amssymb}
\usepackage{graphicx}
\usepackage{tikz-cd}
\usepackage{braket}
\usepackage{slashed}

\newcommand{\dP}{\slashed{p}}
\newcommand{\dk}{\slashed{k}}

\newcommand{\dA}{\slashed{A}}

\newcommand{\bc}{{\mathbf{c}}}

\DeclareMathOperator{\cp}{\slashed \partial}
\DeclareMathOperator{\cA}{\slashed A}
\DeclareMathOperator{\tr}{\text{tr}}
\usetikzlibrary{decorations.pathmorphing}
\usetikzlibrary{decorations.markings}
\usetikzlibrary{math,backgrounds,calc}

\tikzset{
    photon/.style={decorate, decoration={snake}},
    electron/.style={draw=black, postaction={decorate},
        decoration={markings,mark=at position .55 with {\arrow[draw=black]{<}}}},
    relectron/.style={draw=black, postaction={decorate},
        decoration={markings,mark=at position .55 with {\arrow[draw=black]{>}}}}
        }
\tikzstyle{vertex} = [fill,shape=circle,minimum size=3pt,inner sep=2pt, label = below:#1]
\tikzstyle{root} = [draw = none,inner sep = 0pt, label = below:\footnotesize #1]
\tikzstyle{leaf} = [draw = none, inner sep = 0pt , label = \footnotesize #1]
\tikzstyle{rvertex} = [fill,shape=circle,minimum size=3pt,inner sep=2pt, label = right:#1]

\newcommand{\ivo}[1]{\textcolor{red}{#1}}

\begin{document}
\begin{flushright}
	LMU-ASC 29/21\\
	HU-EP-21/24
\end{flushright}

\title{\boldmath QFT with Stubs 
}

\author[a]{Christoph Chiaffrino\note{Corresponding author.}}
\author[b]{and Ivo Sachs}


\affiliation[a]{Institute for Physics, Humboldt University Berlin,
Zum Großen Windkanal 6, D-12489 Berlin, Germany}
\affiliation[b]{Arnold-Sommerfeld-Center for Theoretical Physics, Ludwig-Maximilians-Universit\"at of Munich,\\ Theresienstr. 37, D-80333 M\"unchen, Germany}

\emailAdd{chiaffrc@hu-berlin.de}
\emailAdd{ivo.sachs@physik.lmu.de}


\abstract{The BV-Laplacian $\Delta$ in quantum field theory is singular, by construction, but can be regularized  by deforming the classical BV-action. 
Taking inspiration from string theory we describe a non-local deformation of the latter by adding stubs to the interaction vertices while keeping classical BV-invariance manifest. This is achieved using a version of homotopy transfer resulting in a non-polynomial action for which the quantum master equation is now well defined and will be satisfied by adding additional vertices at loop level. The latter can be defined with the help of standard regularization schemes and is independent of the definition of $\Delta$. In particular, the determination of anomalies reduces to the standard text-book calculation. Finally, we describe how the deformed (quantum) action can be obtained as a canonical transformation. As an example, we illustrate this procedure for quantum electrodynamics.
}

\maketitle
\flushbottom

\section{Introduction}
The BV-formulation of gauge theory is a powerful formalism when one is confronted with open constraint algebras as is the case, for instance, in supergravity or string theory. For standard quantum field theory, while not strictly necessary, this formulation proves nevertheless rather useful for the discussion of renormalizability, or gauge-invariance of the effective action (e.g. \cite{Weinberg:1996kr} and references therein). For instance, in approaches with a momentum cut-off,  where the standard gauge symmetry cannot be maintained in the presence of the cut-off, an embedding into a BV-theory is still possible 
(e.g. \cite{Igarashi:2009tj} and references therein). Also, when a standard quantum field theory is obtained in terms of the BRST-quantization of the world line, the classical BV-extension of the gauge theory arises almost automatically for the same reasons as in string field theory (e.g. \cite{Doubek:2020rbg,Erbin} and references therein).

An important aspect of the quantum BV-master equation concerns anomalies. These are absent if a solution to the BV-master equation can be found. While this condition is easy to formulate, within field theory there does not seem to be a canonical way to construct a regularised BV-action with which the quantum master equation can be tested. See, however, \cite{Barnich:2000me} for general results as well as \cite{Paris:1995dx} for a non-local regularization and \cite{Troost:1989cu} for a Pauli-Villars regularization of the BV-operator. A non-local regularization of BV-Laplacian $\Delta$, compatible with the Wilsonian renormalization group flow was put forward in \cite{Morris:2018axr,Igarashi:2019gkm}. In practical applications, rather than testing the quantum BV-master equation, one usually analyzes possible anomlies by other means, such as the non-conservation of classically conserved currents which can be detected using some standard regularization that may not extend to the full BV action. This is in contrast to string theory where the BV-structure is transferred from the geometric master equation on the moduli space of punctured Riemann surfaces, together with a canonical regularization that consists of adding stubs (short cylinder of length $\ell$) attached to the punctures \cite{Zwiebach:1992ie}. The geometric decomposition of the moduli space produces a solution to the geometric BV-equation that involves infinitely many vertices. This structure is transferred by the world sheet conformal field theory, being a chain map, to the regularized space-time action that solves the quantum BV-equation.

In field theory we do not have an underlying geometric BV-structure (see 
 \cite{Dai:2008bh,Tourkine:2013rda} for attempts), but one can mimic the stubs regularization of string theory by noticing that the latter is realised in space-time by adding a short proper time propagation to the field theory vertices. This will lead to higher order vertices whose form is dictated by the homotopy transfer formula. In fact, in field theory, usually only a small subset of all vertices require regularisation for the BV-operator to be well defined. Thus, unlike regularisation of the geometric punctures, which treats all fields in the same way, in field theory we may add stubs only to the singular vertices which simplifies considerably the construction of the latter. In this note we will provide a solution, to all orders for the stub-regularized classical BV-action for QED and and the fermionic part of QCD. 

The verification of whether such an action satisfies the quantum BV-equation can be done explicitly to all orders in the higher order vertices, or more efficiently by formulating the deformation as a canonical transformation (e.g. \cite{Gomis:1994he,Mnev:2017oko} for a pedagogical review), in which case the quantum BV-equation is automatically satisfied. In fact, canonical transformations are the natural framework to describe deformations of a given BV-action. On the other hand, while BV actions related by canonical transformations formally satisfy the quantum master equation there is no simple way to specify those canonical transformations which regularize the BV-operator. The procedure described in this note is thus to first pick a regularisation which is motivated geometrically and then show that it can be implemented by a canonical transformation. Note that this canonical transformation does not modify the quadratic term of the action, rather it adds stubs to the interaction vertices. 
In addition, the generating function for this canonical transformation generates additional  quantum vertices to the classical action which are necessary to solve the quantum master equation. The proper definition of the latter may or may not lead to anomalies depending on the theory but, importantly, is independent of the definition of $\Delta$ itself. If the theory in question has an anomaly the canonical transformation is still well defined but does not interpolate between solutions of the QME.  As an illustration we will apply this scheme to vector- and axial QED and comment on the generalization to non-abelian theories such as QCD. 

Of course, anomalies have been analyzed within the BV-formalism before, see \cite{Gomis:1994he} and references therein for an overview. The difference between the approach described here and earlier calculations is that we first perform a {\it non-local}\footnote{Or more appropriately, quasi-local.} deformation, $S(\tau)$ of the classical action $S$ so that the action of $\Delta$ is well defined. This step is essentially geometric and independent of the UV-properties of the theory in question. In particular, the compatibility of $\Delta$ and the BV-bracket $\{\,\cdot,\,\cdot\}$ is manifest since these operations are not deformed. Similarly,  a the classical BV-equation is guaranteed by construction to all orders in the deformation parameter. The analysis of the anomalies is then reduced to the identification of (quasi-local) quantum vertices $I_q$ which cancel $\Delta S(\tau)$. In fact these are provided automatically by the generating function for the non-local deformation but need to be normal ordered. This last step can be done with the help of standard methods  independent of the deformation of $S$, which, in particular reduces the question of BV-anomalies to textbook arguments for anomalous Ward-identities. We hope that this helps clarify  the relation between the BV-analysis and the latter approach.

This paper is organized as follows: In section \ref{SP} we describe the regularization of the classical action by adding stubs to the interaction vertices of a scalar field theory. This is conveniently done with the help of Schwinger parameters and can be described without explicit reference to the BV-extension of the theory. In section \ref{BVDS} we review aspects of the BV-formulation and canonical transformations relevant for the following sections. In section \ref{sec:CS} we illustrate the homotopy transfer to the non-local BV-action for chiral QED in 2 dimensions up to fourth order in the fields which is sufficient to reproduce the anomaly of that model. In section \ref{HT} we provide a systematic derivation of the stub regulated action to all orders and show that it satisfies the classical BV-equation. In section \ref{GF} we obtain the exact generating function for the canonical transformation that generates the stub regularization of the classical action and use it to obtain a flow equation for the quantum vertices needed to cancel the BV-laplacian $\Delta S(\tau)$. The formal integration of this equation requires counter terms which may or may not be invariant under classical gauge transformation determining possible anomalies. This step is shown to be equivalent to the calculation of standard text-book triangle amplitudes. In this section we make this explicit for 4-dimension vector QED. In section \ref{aqed} we the repeat this for axial QED which is anomalous.  Finally, we present the conclusions.

\section{Schwinger Parametrization}\label{SP}

In order to explain the concept of stubs in field theory we first consider the Klein-Gordon operator $\square + m^2$,\footnote{We use the convention $\eta = \text{diag}(1,-1,...,-1)$.} whose inverse $P(\infty)$, describes the propagation of a free scalar field. It can be formally written as
\begin{equation}
P(\infty) := \frac{1}{\square + m^2} = \int_0^\infty \text d s \, e^{-s (\square + m^2)} \, .
\end{equation}
This is just the familiar Schwinger parametrization. The argument of $P$ is there to indicate that we integrate to infinity. Some care should be taken in taking the integral  limit to infinity. However, we will see that the following argument does not rely on this detail. The integral representation is only there to highlight that the propagator can be written as a sum of two pieces.
\begin{equation}\label{PropSplit}
\begin{split}
\int_0^\infty \text d s \, e^{-s (\square + m^2)} &= \int_0^\tau \text d s \, e^{-s (\square + m^2)} + \int_\tau^\infty \text d s \, e^{-s (\square + m^2)}  \\
&= P(\tau) + K(\tau )\circ P(\infty) \, ,
\end{split}
\end{equation}
where $K(\tau)$ for the heat kernel of $\square + m^2$, i.e.
\begin{equation}
K(\tau) = e^{-\tau (\square + m^2)} \, .
\end{equation}
The crucial observation then, is that
\begin{equation}\label{spk}
(\square + m^2) P(\tau) = 1 - K(\tau) \, .
\end{equation}
This is well defined without specifying how we define $P(\infty)$ and is a manifestation of the fact that $K(\tau)$ is homotopic to the identity. This is the property we need for what follows. 

Now we consider an interacting theory. For concreteness we assume that there is a cubic interaction $S_I := \int \frac{\lambda}{3!}\phi^3$. Our goal is then to find a physically equivalent interaction based on the homotopy \eqref{spk}. The physical observables of the theory are the scattering amplitudes.  At tree level (i.e. to lowest order in $\hbar$), the amplitude for four wave packets $\phi_i, i = 1,...,4$ to scatter is
\begin{align}\label{4pt}
\mathcal A(\phi_1,\phi_2,\phi_3,\phi_4) = \lambda^2\int \text d^n x \, \text d^n y \,  \phi_1(x)\phi_2(x)P_\infty(x,y) \phi_3(y)\phi_4(y) \;+  \; (t,u)\,\text{channels} ,
\end{align}
where $P_\infty(x,y)$ is the integral kernel of the propagator $P(\infty)$ and fields $\phi_i$ satisfy the free equations of motion. Using \eqref{PropSplit}, we can write equivalently, 
\begin{equation}
\begin{split}\label{4ptSplit}
\mathcal A(\phi_1,\phi_2,\phi_3,\phi_4) =& \lambda^2\int \text d^n x \, \text d^n y \,  \phi_1(x)\phi_2(x) P_\tau(x,y)  \phi_3(y)\phi_4(y)  \\
&+ \lambda^2 \int \text d^n x \, \text d^n y \, \text d^2 z \, \phi_1(x)\phi_2(x) K_\tau(x,y) P_\infty(y,z) \phi_3(z) \phi_4(z)  \,  ,
\end{split}
\end{equation}
where $K_\tau(x,y)$ and $P_\tau(x,y)$ denote the integral kernels of $K(\tau)$ and $P(\tau)$ respectively. We emphasize that while \eqref{4ptSplit} equals \eqref{4pt}, this rewriting suggests a different interaction reproducing this amplitude. The second contribution to \eqref{4ptSplit} again involves the propagator $P(\infty)$, suggesting that this piece comes from some cubic interaction. The first term, on the other hand, has no propagator, so we may attribute it to a quartic interaction. In fact, an equivalent interaction reproducing \eqref{4ptSplit} (and therefore also \eqref{4pt}) is given by
\begin{equation}\label{effectivescalartheory}
S_I(\tau) = \tfrac{\lambda}{3!}\int \text d^n x \, [K(\tfrac{\tau}{2})\phi(x)]^3 + \tfrac{\lambda^2}{4!} \int \text d^n x \, \text d^n y \, [K(\tfrac{\tau}{2})\phi(x)]^2 P_\tau(x,y) [K(\tfrac{\tau}{2})\phi(y)]^2 \, .
\end{equation}
The four point tree-level scattering derived from the cubic interaction is
\begin{equation}\label{effectivecubic4pt1}
\begin{split}
\mathcal{A}_1(\phi_1,...,\phi_4) = \lambda^2\int & \text d^n w \, \text d^n x \, \text d^n y \, \text d^n z \, [K(\tfrac{\tau}{2})\phi_1(x)][K(\tfrac{\tau}{2})\phi_2(x)] \\
\times & K_{\frac{\tau}{2}}(x,y)P_\infty(y,z) K_{\frac{\tau}{2}}(z,w)[K(\tfrac{\tau}{2})\phi_3(w)][K(\tfrac{\tau}{2})\phi_4(w)] \, .
\end{split}
\end{equation}
Using $K(\tau) \circ P(\infty) = P(\infty) \circ K(\tau)$ and $K(\tau_1)\circ K(\tau_2) = K(\tau_1 + \tau_2)$, \eqref{effectivecubic4pt1} reduces to
\begin{equation} \label{effectivecubic4pt2}
\begin{split}
\mathcal{A}_1(\phi_1,...,\phi_4) = \lambda^2\int & \text d^n x \, \text d^n y \, \text d^n z \, [K(\tfrac{\tau}{2})\phi_1(x)][K(\tfrac{\tau}{2})\phi_2(x)] \\
\times & K_{\tau}(x,y)P_\infty(y,z)[K(\tfrac{\tau}{2})\phi_3(z)][K(\tfrac{\tau}{2})\phi_4(z)] \, .
\end{split}
\end{equation}
The $\phi_i$ are on shell, so we have $K(\tfrac{\tau}{2})\phi_i = \phi_i$. Therefore, \eqref{effectivecubic4pt2} immediately reduces to the second term in \eqref{4ptSplit}. By the same reasoning, the quartic interaction of \eqref{effectivescalartheory} reproduces the first term in \eqref{4ptSplit}.

We just learned that an interaction of the form \eqref{effectivescalartheory} provides the same four point scattering amplitude as a $\phi^3$ theory (at tree level). To obtain a physically equivalent theory, we should check that the amplitudes agree for any number of loops and particles. In order for this to be the case, in addition to the quartic term, we need to introduce even higher interactions. Rather than doing this order by order, it would be advantageous to have a formalism that tells us how to proceed in general. We will see that such a procedure can be derived using the language of the Batalin-Vilkovisky (BV) formalism and the homotopy transfer for the associated $L_\infty$-algebras.

\section{Review of the BV Formalism} \label{BVDS}

We first review some aspects of the BV formalism (e.g. \cite{Gomis:1994he, Weinberg:1996kr} for a review), which is based on a $\mathbb{Z}-$graded vector space $X^\bullet = \sum_{n \in \mathbb{Z}} X^n$ of fields\footnote{In its most general form, the BV language deals with $\mathbb{Z}$-graded manifolds. For our purposes it will be enough to only consider linear spaces.}. The grading on $X^\bullet$ is given by the ghost number. The fields in $X^0$ (ghost number zero) are those we are familiar with, from the standard formulation of quantum field theories. In case of the scalar field theory considered in the last section, $X^0$ would consist of scalar fields $\phi$, while for Yang-Mills theory the fields in $X^0$ are gauge potentials $A_\mu$. When coupling Yang-Mills to fermions, we also include spin $\frac{1}{2}$ fields $\psi$. The space $X^{1}$ contains the BRST ghosts. For a scalar theory we would take $X^{1} = \{0\}$, while in Yang-Mills (with or without fermions) we have a set of ghosts $c^a(x),a = 1,...,N$, where $N$ is the dimension of the gauge group. In this work we only encounter irreducible gauge theories, in which case $X^{n \ge 2} = \{0\}$. 
In negative ghost number we essentially have $X^{-k} \cong X^{k-1}, k \ge 1$, but with reversed parity. This means that, for example, when a field $\phi \in X^{k - 1}$ is bosonic, its corresponding field $\phi^* \in X^{-k}$ is fermionic. Fields in $X^{-1}$ are referred to as anti-fields, while those in $X^{k \le -2}$ are collectively referred to as anti-ghosts but often also just anti-fields. In particular, given any field $\Phi^A \in X^{n \ge 0}$, we call $\Phi^*_A \in X^{-n-1}$ a conjugate anti-field. The index $A$ collectively stands for all indices that $\Phi$ may have, e.g. spacetime, spin or internal. We denote by $\epsilon(\Phi) \in \mathbb{Z}/2\mathbb{Z}$ for the statistics of any object $\Phi \in X^\bullet$. All this is summarized in the following table:
\begin{center}
\begin{tabular}{ |c|c|c|c| } \hline
Space & Name & Notation & Statistics\\ \hline
$\vdots$ & $\vdots$ & $\vdots$ & $\vdots$ \\
$X^{1}$ & ghosts & $c^a$ & fermionic  \\
$X^0$ & fields & $\Phi = \phi,A_\mu,\psi, h_{\mu\nu},...$ & $\epsilon(\Phi)$ \\ 
$X^{-1}$ & anti-fields & $\Phi^* = \phi^*,A_\mu^*,\psi^*, h_{\mu\nu}^*,...$ & $\epsilon(\Phi)$ + 1 \\
$X^{-2}$ & anti-ghosts & $c_a^*$ & bosonic\\
$\vdots$ & $\vdots$ & $\vdots$ & $\vdots$ \\ \hline
\end{tabular}\label{bva1}
\end{center}

The BV-bracket is a graded Poisson bracket, realized on the space of functionals on $X^\bullet$ as 
\begin{equation}\label{bvb}
\{F,G\} = \sum_{Y} \int \text d^n x \,  \frac{\delta_r F}{\delta \Phi_A^*(x)}\frac{\delta_l G}{\delta \Phi^A(x)} - \frac{\delta_r F}{\delta \Phi^A(x)}\frac{\delta_l G}{\delta \Phi_A^*(x)} \, ,
\end{equation}
where the subscripts $l,r$ on the functional derivatives indicate that one should apply Leibniz' rule from the left/right. The BV-bracket is graded symmetric,
\begin{equation}
\{F,G\} = - (-)^{(\epsilon(F)+1)(\epsilon(G)+1)}\{G,F\} \, 
\end{equation}
and satisfies the graded Jacobi identity
\begin{equation}
(-)^{(\epsilon(F)+1)(\epsilon(H)+1)} \{F,\{G,H\}\} + \text{cyclic} = 0 \, . 
\end{equation}
Finally, this bracket is compatible with the 
BV-Laplacian
\begin{equation}\label{BVD}
\Delta(F) = \sum_\Phi \int \text d^n x \, (-)^{\epsilon(\Phi)+1}\frac{\delta_l^2 F}{\delta \Phi^A(x) \delta \Phi_A^*(x)} \,,\qquad \Delta^2=0\,, 
\end{equation}
that is
\begin{equation}\label{LaplacePoisson}
\begin{split}
\Delta(FG) &= \Delta(F)G + (-)^{\epsilon(F)} \{F,G\} + (-)^{\epsilon(F)} F \Delta(G) \, , \\
\Delta \{F,G\} &= \{\Delta(F),G\} + (-)^{\epsilon(F) + 1}\{F,\Delta{G}\} \, .
\end{split}
\end{equation}
We should note that the definition \eqref{LaplacePoisson} is a formal one since it involves two functional derivations at the same point. One way to define $\Delta$ is to restrict the space of functionals on which $\Delta$ acts. 

\subsection{Classical BV action}
For a given classical gauge theory there is a simple procedure to construct its BV-extension as\footnote{While there are more general BV-actions that are not obtained in this way we will not need them here.} 
\begin{align}\label{BVs}
    S[\Phi,\Phi^*]=S[\Phi]+\int s(\Phi^A)\Phi_A^*
\end{align}
where $S[\Phi]$ is the classical  action completed by the ghost degrees of freedom needed for the BRST-quantization procedure and $s(\Phi^A)$ is the BRST-transformation (BRST-vector field) for the latter. The action (\ref{BVs}) then satisfies the classical BV-master equation 
\begin{align}
    \{S,S\}&=2\sum\limits_I\int \frac{\delta_r S}{\delta \Phi_A^{*}}\frac{\delta_l S}{\delta \Phi^I}=0\,.
\end{align}
In the path-integral quantization of (\ref{BVs}) one expresses the anti-fields in terms of the fields by means a gauge fixing fermion $\Psi(X)$ as 
\begin{align}
    \int \mathcal{D}(\Phi, \Phi^*)\;\delta(\Phi_A^*-\frac{\delta_l\Psi}{\delta \Phi_A})\;e^{\frac{i}{\hbar}S[\Phi,\Phi^*]}\,,
\end{align}
where $\delta(\Phi^*-\cdots)$ is a functional Dirac delta-function. The invariance of this procedure under different choices of $\Psi$ translates into the condition 
\begin{align}
  0=  \int \mathcal{D}(\Phi)\;F(\Phi,\Phi^*)\sum\limits_A\int\left(
 \frac{\delta_r S}{\delta \Phi_A^{*}}\frac{\delta_l S}{\delta \Phi^A}-2i\hbar\frac{\delta_l \delta_r S}{\delta \Phi^A\delta \Phi_A^{*}}\right)\;e^{\frac{i}{\hbar}S[\Phi,\Phi^*]}|_{\Phi_A^*=\frac{\delta_l\Psi}{\delta \Phi^A}}\,,
\end{align}  
for all gauge-invariant functionals, $F$. This is recognized as the quantum BV-equation with the $\hbar$ term being twice the $BV$-operator $\Delta$. 
Thus, a quantum consistent action $S$ \ivo{$S_q$ ?} satisfies the quantum master equation
\begin{equation}\label{QME}
i \hbar \Delta(S) - \frac{1}{2}\{S,S\} = 0 \, . 
\end{equation}
Using the Jacobi identity of $\{-,-\}$ together with the properties listed in \eqref{LaplacePoisson}, one can show that this is equivalent to the condition that the operator
\begin{equation}
Q_q = \{S,-\} - i\hbar \Delta
\end{equation}
squares to zero.
The action $S$ may have quantum corrections that come with powers of $\hbar$. We define the classical action $S|_{\hbar = 0}$. One usually assumes that \eqref{QME} is satisfied in each order of $\hbar$. If that is the case, the quantum master equation then implies the classical master equation
\begin{equation}\label{CME}
\{S,S\} = 0
\end{equation}
by considering \eqref{QME} to zeroth order in $\hbar$. Similarly,
\begin{equation}\label{qcl1}
Q_{cl} = \{S,-\} \, ,
\end{equation}
evaluated at $\hbar=0$. 
The observables of the theory live in the cohomology $H = \frac{\ker Q_q}{\text{Im} \, Q_q}$ of $Q_q$.

Given an action $S$ satisfying the quantum master equation, any deformation of the form
\begin{equation}\label{GenFun}
S \mapsto S +\delta S := S + \tau \{S,R\} - \tau i\hbar \Delta(R)
\end{equation}
also solves the quantum master equation to linear order in $\tau$. The functional $R$ is called a generating functional for this transformation. Note that for \eqref{GenFun} to make sense, $R$ has to be odd and of ghost number minus one. We can use the same generating functional to transform functionals $F$ of the fields as 
\begin{equation}
\delta F = \tau \{F,R\} - \tau i \hbar \Delta(R) \, .
\end{equation}
Such a transformation preserves the cohomology. We therefore can consider two theories connected by a generating functional to be equivalent. 

To conclude this section we note that the operator $Q_{cl}$ in \eqref{qcl1} can be expanded around any given background $\bar\Phi$ in ghost degree zero. For this we write 
\begin{align}
    Q_{cl}=Q^{(0)}+Q^{(1)}+Q^{(2)}+Q^{(3)}+\cdots\,,
\end{align}
where $(k)$ denotes the power in the fields. Recalling \eqref{bvb} we see that $Q^{(0)}$ vanishes provided the background satisfies the classical equation of motion, which we will assume at present, while $Q^{(1)}$ has the interpretation as a linear map: $T_{\bar\Phi}\to T_{\bar\Phi}$ from the tangent space at $\bar\Phi$ to itself. Similarly, $Q^{(k)}$ defines a map 
\begin{align}\label{qexp}
    Q^{(k)}:\quad T_{\bar\Phi}^{\otimes k}\to T_{\bar\Phi}\,.
\end{align}
Nilpotency of $Q_{cl}$ then induces on $\{Q^{(k)}\}$ the structure of an $L_\infty$-algebra. Some of its features will be reviewed in in section \ref{HT}. For now let us just note that the vertices of the classical action, together with the BV-bracket \eqref{bvb} have the structure of a homotopy Lie algebra. Since this holds for the initial BV-action as well as for he BV-action constructed form the deformation described in section \ref{SP}, this deformation can be implemented as a homotopy transfer from one $L_\infty$-algebra to another. We will illustrate this in the next section for 2-dimensional $QED$ to lowest order and then describe the complete quantum action in the subsequent sections.

\section{Chiral $QED_2$}
\label{sec:CS}
Let us first illustrate the stubs regularization for 2-dimensional, chiral $QED$ which is known to be anomalous. For this we first recall its action
\begin{equation}\label{chiralS}
S_0 = \int \sqrt{2} \bar{\psi}(i\partial_- + A_-) \psi -\frac{1}{4}F_{\mu\nu}F^{\mu\nu}.
\end{equation}
The field $\bar{\psi}$ denotes the complex conjugate of $\psi$. We use lightcone coordinates $x^\pm = \frac{1}{\sqrt{2}}(t \pm x)$. The BV-extension is obtained by adding the term
\begin{equation}\label{SS1}
S_1 = \int ic \bar{\psi}^* \bar{\psi} - i c \psi^* \psi - A_\mu^* \partial^\mu c.
\end{equation}
As it stands, for this classical BV-action of the BV Laplacian is not defined on $S_1$ since for $\Phi=\psi$ or $\bar\psi$ the Laplacian  $\Delta$ \eqref{BVD} is singular. We can fix this by considering a new action $S(\tau)$, which is equivalent to $S=S_0 + S_1$ at the classical level. It can be constructed with the help of a homotopy with respect to the linear differential $Q^{(1)}$ generated by $S$ at $\bar\Phi=0$ and which interpolates between the identity and the stub $K$ as in \eqref{spk}. After identifying  $Q_0\equiv Q^{(1)}$ in \eqref{qexp} we have
\begin{equation}
Q_0 = \{S^{(2)},\, \cdot \, \} = \int_{x}\sqrt{2}i\partial_- \psi(x) \frac{\delta}{\delta \bar{\psi}^*(x)} + \sqrt{2}i\partial_- \bar{\psi}(x)\frac{\delta}{\delta \psi^*(x)} + ... \, ,
\end{equation}
where we omitted everything that only involves the gauge field and the ghost, since they will not be important for us. The reason is that actions of the BV-Laplacian vanishes identically, except for its action on fermions. Following the proposal in section \ref{SP} we regularize the latter by the replacement\footnote{Below we will sometimes suppress the label $\tau$ when there is no risk of confusion.}
\begin{equation}
\psi \mapsto K\psi := e^{-\tau  \square} \psi,
\end{equation}
and similarly for the complex conjugate and all the anti-fields. This can be justified by noting that $e^{-\tau  \square}$ is homotopic to the identity operator. The corresponding homotopy is
\begin{equation}
H = -i\sqrt{2}(\int_x \int_0^{\tau } \bar{\psi}^*(x) e^{-\tau\square}\partial_+ \frac{\delta}{\delta \psi(x)} + \psi^*(x) e^{-\tau\square}\partial_+ \frac{\delta}{\delta \bar{\psi}(x)}).
\end{equation}
One then computes
\begin{equation}
\{Q_0,H\} = \int_x \psi(x)\frac{\delta}{\delta \psi(x)} - \int_x \psi(x)e^{-\tau \square}\frac{\delta}{\delta \psi(x)} + ... \,
\end{equation}
where the dots contain three more contributions which are obtained by replacing $\psi \mapsto (\psi^*,\bar{\psi},\bar{\psi}^*)$. The operator $\int_x \psi(x)\frac{\delta}{\delta \psi(x)}$ is recognized as the identity operator on field space, while $\int_x \psi(x)e^{-\tau \square}\frac{\delta}{\delta \psi(x)}$ is, of course, the heat kernel. In other words, 
\begin{align}
    \{Q_0,H\} &=1-K
\end{align}
and $K$ is indeed homotopic to the identity operator. In order to see how this homotopy is related to $P(\tau)$ and $K(\tau)$ in section \ref{SP} we write
\begin{equation}\label{p2}
P(x,y) = -i\sqrt{2} \int_0^{\tau }e^{-\tau \square}\partial_+ \delta(x-y)\qquad\text{and}\quad
K(x,y) = e^{-\tau  \square}\delta(x-y).
\end{equation}
Then,
\begin{equation}\label{PropagatorRelation}
\sqrt{2}i\partial_{x^-}P(x,y) = \delta(x-y) - K(x,y).
\end{equation}

Returning to the BV-formalism we would like to interpret $K(\tau)$ in terms of a homotopy transfer between the vertices of the classical BV-action and its deformation induced by the above regularization. For this denote by $V$ the vector space of the spinor fields (as well as their conjugates and anti fields). On $V$ we define  the homotopy equivalence data by
\begin{equation}
i: (V,Q_0) \leftrightarrows (V,Q_0) : p.
\end{equation}
There is some freedom in choosing the maps $p$ and $i$. All we need for a homotopy transfer is that $i\circ p = K$. We make a symmetric choice. For this we define
\begin{equation}\label{khp}
K_{\frac{1}{2}}\psi := e^{-\frac{\tau }{2}\square}\psi
\end{equation}
and set $p = i = K_{\frac{1}{2}}$. Furthermore, we denote by $K_{\frac{1}{2}}(x,y)$ the integral kernel of $K_{\frac{1}{2}}$.

As explained at the end of section \ref{BVD}  $Q_{cl}=\{S, \, \cdot \, \}$ defines an $L_\infty$-algebra. The homotopy transfer then provides a new $L_\infty$-structure induced by the regularized action $S(\tau)$ and related to the original one by the homotopy $H$. It is obtained by constructing Feynman diagrams with propagator $H$ and external legs (stubs) $K_{\frac{1}{2}}$. To cubic order, the new action $S(\tau)$ is simply given by inserting the stubs for the fermions and their antifields. At cubic order we have
\begin{align}
S_3(\tau) = \sqrt{2}\bar{\psi}(z)K_{\frac{1}{2}}(z,y)A_-(y)K_{\frac{1}{2}}(y,x)\psi(x) & \nonumber \\
- i\psi^*(z)K_{\frac{1}{2}}(z,y)c(y)K_{\frac{1}{2}}(y,x)\psi(x)  & \\
- i \bar{\psi}(z)K_{\frac{1}{2}}(z,y)c(y)K_{\frac{1}{2}}(y,x)\bar{\psi}^*(x) & \nonumber \, .
\end{align}
The above expression uses Einstein - de Witt integration convention, which we will use from now on. If we had a homotopy also affecting the ghost and gauge field, then there would be a similar stub contribution for these fields. Unless $K_{\frac{1}{2}} = 1$, the above expression gives a non-vanishing contribution to the master equation,
\begin{align}
\{S_3(\tau),S_3(\tau)\} = -\sqrt{8}i\bar{\psi}(u)K_{\frac{1}{2}}(u,w)c(w)K(w,y)A_-(y)K_{\frac{1}{2}}(y,x)\psi(x)  &\nonumber\\
+\sqrt{8}i\bar{\psi}(u)K_{\frac{1}{2}}(u,w)A_-(w)K(w,y)c(y)K_{\frac{1}{2}}(y,x)\psi(x) &\nonumber \\
+2 \psi^*(u)K_{\frac{1}{2}}(u,w)c(w)K(w,y)c(y)K_{\frac{1}{2}}(y,x)\psi(x) & \\
+2 \bar{\psi}^*(u)K_{\frac{1}{2}}(u,w)c(w)K(w,y)c(y)K_{\frac{1}{2}}(y,x)\bar{\psi}(x) & .\nonumber
\end{align}
The induced quartic interaction exactly cancels this. It is given by
\begin{align}
S_4(\tau) =& -\left(\sqrt{2}\bar{\psi}(u) K_{\frac{1}{2}}(u,w) A_-(w) + i\psi^*(u)K_{\frac{1}{2}}(u,w)c(w))\right) \\
&\cdot P(w,y)\left(\sqrt{2}A_-(y)K_{\frac{1}{2}}(y,x)\psi(x)+ ic(y)K_{\frac{1}{2}}(y,x)\bar{\psi}^*(x)\right).\nonumber
\end{align}
In order to see this we may use the identities 
\begin{align}
Q_0(\sqrt{2}\bar{\psi}(u) K_{\frac{1}{2}}(u,w) A_-(w) + i\psi^*(u)K_{\frac{1}{2}}(u,w)c(w)) \\ 
= - \sqrt{2}\partial_{w^-}(\bar{\psi}(u)K_{\frac{1}{2}}(u,w)c(w)),\nonumber
\end{align}
and
\begin{align}
Q_0(\sqrt{2}A_-(y)K_{\frac{1}{2}}(y,x)\psi(x)+ ic(y)K_{\frac{1}{2}}(y,x)\bar{\psi}^*(x)) \\
 = \sqrt{2}\partial_{y^-}(c(y)K_{\frac{1}{2}}(y,x)\psi(x)).\nonumber
\end{align}
It then follows that
\begin{align}
Q_0 S_4(\tau) = - \sqrt{2}\bar{\psi}(u)K_{\frac{1}{2}}(u,w)c(w)\partial_{w^-}P(w,y)(\sqrt{2}A_-(y)K_{\frac{1}{2}}(y,x)\psi(x)\nonumber \\
+ ic(y)K_{\frac{1}{2}}(y,x)\bar{\psi}^*(x)) -\sqrt{2} (\sqrt{2}\bar{\psi}(u) K_{\frac{1}{2}}(u,w) A_-(w)  \\ 
+ i\psi^*(u)K_{\frac{1}{2}}(u,w)c(w))\partial_{y^-}P(w,y)c(y)K_{\frac{1}{2}}(y,x)\psi(x)\nonumber
\end{align}
Recalling (\ref{PropagatorRelation}) this becomes
\begin{align}
Q_0 S_4(\tau)	= -i \bar{\psi}(u)K_{\frac{1}{2}}(u,w)c(w)K(w,y)(\sqrt{2}A_-(y)K_{\frac{1}{2}}(y,x)\psi(x)+ \nonumber\\ ic(y)K_{\frac{1}{2}}(y,x)\bar{\psi}^*(x))
				+ i (\sqrt{2}\bar{\psi}(u) K_{\frac{1}{2}}(u,w) A_-(w) \\
				+ i\psi^*(u)K_{\frac{1}{2}}(u,w)c(w))K(w,y)c(y)K_{\frac{1}{2}}(y,x)\psi(x).\nonumber
\end{align}
From this we then easily see that 
\begin{equation}
2 Q_0 S_3(\tau) + \{S_3(\tau),S_3(\tau)\} = 0.
\end{equation}
Hence, the quartic vertex exactly cancels the contribution from the cubic vertex, as it should according to the homotopy transfer theorem.

We are now in position to verify the quantum BV-equation since, on $S(\tau)$, the BV-Laplacian is well defined. Acting with $\Delta$ on $S(\tau)$ we have for  $\Delta S(\tau)$ in cubic and quartic order, 
\begin{equation}\label{ds3}
\Delta S_3(\tau) = - i K_{\frac{1}{2}}(x,y)c(y)K_{\frac{1}{2}}(y,x) + i K_{\frac{1}{2}}(x,y)c(y)K_{\frac{1}{2}}(y,x) = 0
\end{equation}
and
\begin{align}\label{ds4}
\Delta S_4(\tau) =& -i\sqrt{2}K_{\frac{1}{2}}(x,w)A_-(w)P(w,y) c(y)K_{\frac{1}{2}}(y,x)  \\ &+ i \sqrt{2}K_{\frac{1}{2}}(x,w)c(w)P(w,y)A_-(y)K_{\frac{1}{2}}(y,x) 
 = i\sqrt{8}K(y,w)c(w)P(w,y)A_-(y).\nonumber
\end{align}
We used that $K_{\frac{1}{2}}(x,y)K_{\frac{1}{2}}(x,z) = K(x,z)$ and  $P(x,y) = - P(y,x)$. The result  $\tau $-dependent, which means that $S(\tau)$ has to be further completed with the addition of of $\hbar$- and $\tau $ dependent quantum vertices. We will return to this in section \ref{CT}. For now let us just focus on the $\tau$-independent contribution. For this we take the limit $\tau  \rightarrow 0$. Using 
\begin{equation}
P(x,y) = \sqrt{2}i\tau \partial_{x^+} \delta(x-y) + \mathcal{O}(\tau ^2)\,,
\end{equation} 
and
\begin{equation}
K(x,y) = \frac{1}{4\pi i \tau } e^{-\frac{(x-y)^2}{4\tau }}\,,
\end{equation}
it follows that
\begin{equation}
\Delta S_4(\tau) = \frac{i}{\pi}\int_x c(x)\partial_+ A_-(x)\,,
\end{equation}
thus reproducing the well known anomaly of chiral $QED_2$. To conclude, 
\begin{equation}
i \hbar \Delta(S(\tau)) - \frac{1}{2}\{S(\tau),S(\tau)\} \neq 0 \, ,
\end{equation}
at 0-th order in the regularization parameter $\tau $. Now, the fact that $S(\tau)$ does not satisfy the quantum BV-equation this does not exclude, per se that one can not add a quantum correction $I_q$ to $S(\tau)$ that cures this deficiency. However, the only Lorentz-invariant correction that is quadratic in the gauge field and of zeroth order in derivatives, i.e. 
\begin{align}
    I_2=\int A_-A_+\,,\quad\text{with}\quad \{Q_0I_2\}= \int \partial_-c A_++A_-\partial_+c
\end{align}
cannot cancel this anomaly\footnote{The  {\it non-local} interaction
$I_{nl} = \frac{\hbar}{2\pi}\int_x A_-(x)\frac{\partial_+}{\partial_-}A_-(x)$
would cancel the anomaly.}.

While this is enough to show that the theory in question is anomalous in this case one may still wonder what the higher order corrections to the regularized BV action $S(\tau)$ are. In general, a solution to that quantum BV equation \eqref{QME} receives corrections to all orders in the deformations parameter as well as quantum corrections in powers of $\hbar$. In the following sections we will give a construction of these higher order terms using homotopy transfer and later describe them in terms of canonical transformations w.r.t. to the BV-bracket.  In section \ref{HT} we then give a systematic derivation of the classical action with stubs to all orders and show that the classical BV-equation is satisfied.  

At this point we would like to mention that the heat kernel as a tool to regularize the BV Laplacian was also used in \cite{costellorenormalization}. In this reference, the heat kernel enters in the definition of the BV Laplacian and the BV bracket. The regularized $\Delta$ is then defined on functionals on which the usual Laplacian would be singular. Since $\{-,-\}$ is also altered\footnote{When regularizing $\Delta$ we also have to change the bracket in order to preserve the relations given in \eqref{LaplacePoisson}.}, the standard BV actions like \eqref{chiralS} no longer satisfy the classical master equation, and one has to add higher order interaction to restore classical BRST invariance even at the classical level. This is very similar to what happens in our approach.

\section{Stubs through Homotopy Transfer in Quantum Electrodynamics}\label{HT}

The classical BV-action of massless quantum electrodynamics in any dimension is given by 
\begin{equation}\label{QEDaction}
S  = \int i \bar{\psi}\gamma^\mu \partial_\mu \psi - \frac{1}{4}F_{\mu\nu}F^{\mu\nu} + A_\mu \bar{\psi}\gamma^\mu \psi - A_\mu^* \partial^\mu c + i \psi^* c \psi + i  \bar{\psi} c \bar{\psi}^* \, ,
\end{equation}
where the first three terms generate the equations of motion, while the last three generate gauge transformations. This action satisfies the classical master equation $\{S,S\} = 0$. We have already seen in the previous section that since $\Delta S$ is singular, the quantum master equation is not defined. As we did there, we deform the classical action \eqref{QEDaction} by introducing stubs for the fermion fields. This is sufficient since the singularity is only due to the part in $S$ that generates the gauge transformation of the fermions. Proceeding as in section \ref{sec:CS}, we modify the vertices by propagating the fermions by the proper time $\frac{\tau}{2}$. The heat kernel $K_{\frac{1}{2}} = e^{-\frac{\tau}{2}\square}$ generates this propagation. Hence, in each vertex we want to make the replacement
\begin{equation}\label{fermionstubs}
(\psi,\bar{\psi},\psi^*,\bar{\psi}^*) \mapsto (K_{\frac{1}{2}}\psi,\bar{\psi}K_{\frac{1}{2}},\psi^*K_{\frac{1}{2}},K_{\frac{1}{2}} \bar{\psi}^*) \, .
\end{equation}
We already saw that this deformation requires us to introduce higher order vertices in order to satisfy the classical master equation. To determine the exact deformation to all orders we can use the homotopy transfer theorem. If we are able to find a homtopy that generates the transformation \eqref{fermionstubs}, the homotopy transfer theorem will tell us how we need to deform equations of motion and gauge transformations such that the theory with stubs is again an $L_\infty$ algebra. That leaves us with the task to write down a BV-action whose vertices generate this $L_\infty$ algebra.

To derive the $L_\infty$ products from \eqref{QEDaction}, we note that the action of the linearization, $Q_{0}$, of $Q_{cl}=\{S ,-\}$ is given by 
\begin{align}
Q_{0}\psi = 0 ,& &Q_{0}\bar{\psi} = 0,& &Q_{0}A^\mu = \partial^\mu c,& & Q_{0}c = 0,&\label{Q0} \\
Q_{0}\psi^* = i \partial_\mu \bar{\psi}\gamma^\mu, & &Q_{0} \bar{\psi}^* = i \gamma^\mu \partial_\mu \psi, &  &Q_{0}A^*_\nu = -\partial^\mu F_{\mu\nu}, & &Q_{0}c^* = - \partial^\mu A_\mu^* .&\nonumber
\end{align}
Comparing this to \eqref{qexp}, we read off the linear $L_\infty$ product $b_1\equiv Q^{(1)}=Q_0$. A general field is a list
\begin{equation}
Y = (c,A^\mu,\psi,\bar{\psi},A_\nu^*,\bar{\psi}^*,\psi^*, c^*) \, ,
\end{equation}
on which the linear map $b_1$ thus acts as
\begin{equation}
b_1 Y = (0,\partial^\mu c,0,0,-\partial^\mu F_{\mu\nu},i \gamma^\mu \partial_\mu \psi,i\partial_\mu \bar{\psi}\gamma^\mu, - \partial^\mu A^*_\mu) \, .
\end{equation}
The interactions determine the the quadratic product.
\begin{equation}
\frac{1}{2}b_2(Y,Y) = (0,0,-ic\psi,-i\bar{\psi} c,\bar{\psi}\gamma_\nu \psi,-\cA \psi - i c\bar{\psi}^*, \bar{\psi}\cA + i \psi^* c, -i\psi^*\psi + i \bar{\psi}\bar{\psi}^*) \, .
\end{equation}
Note that we wrote $b_2$ for identical fields (inputs) in order to save some writing. In general it takes two different inputs. The higher products vanish identically, i.
e.$b_{n \ge 3} = 0$, since we are dealing with a cubic theory.

Next, we want to find a homotopy $h$ such that $b_1 h + h b_1 = 1 - i p$. We make a symmetric choice such that both $i_1$ and $p_1$ generate a fermionic stub of length $\frac{\tau}{2}$. We set
\begin{equation}
i_1(Y) = p_1(Y) = (c,A^\mu,K_{\frac{1}{2}}\psi,K_{\frac{1}{2}}\bar{\psi},A_\nu^*,K_{\frac{1}{2}}\bar{\psi}^*,K_{\frac{1}{2}}\psi^*, c^*) \, .
\end{equation}
Since $[i \cp, K_{\frac{1}{2}}] = 0$, it follows that $[i_1,b_1] = [p_1,b_1] = 0$, so both $i_1$ and $p_1$ are chain maps with respect to $b_1$. As $i_1$ and $p_1$ act as the identity on the fields $(A^\mu,A_\nu^*)$ and ghosts $(c,c^*)$, on these fields we need that $b_1 h + h b_1 = 0$. Therefore, we can just take $h = 0$ for the latter. On the other hand, for each fermion field $(\psi,\bar{\psi},\bar{\psi}^*,\psi^*)$ we have
\begin{equation}
1 - i_1 p_1 = 1 - K_{\frac{1}{2}}\circ K_{\frac{1}{2}} =: 1 - K \, ,
\end{equation}
where, as before,  $K = e^{-\tau \square}$. Then 
\begin{align}\label{PA}
    P = -i \cp \int_0^\tau \text d s \,  e^{-s \square} \,\quad \text{with}\quad i \cp P = P i \cp = 1 - K\,.
\end{align} 
The Ansatz for the homotopy is then
\begin{equation}
h Y = (0,0,P\bar{\psi}^*,P \psi^*,0,0,0,0) \, ,
\end{equation}
from which it follows that
\begin{align*}
h b_1 Y &= (0,0,(1-K)\psi, (1-K)\bar{\psi},0,0,0,0) \, , \\
b_1 h Y &= (0,0,0,0,(1-K)\bar{\psi}^*,(1-K)\psi^*,0,0) \, ,
\end{align*} 
and therefore
\begin{equation}
(h b_1 + b_1 h) Y = (0,0,(1-K)\psi, (1-K)\bar{\psi},(1-K)\bar{\psi}^*,(1-K)\psi^*,0,0) \, ,
\end{equation}
as required for $K$ to be homotopic to the identity.

We are now ready describe the $L_\infty$ algebra induced by the homotopy $h$. Without going into full detail on how the $n$-products $\tilde b_n$ are constructed (e.g. \cite{Doubek:2020rbg} and references therein) we simply use the fact that the $n+1$ particle vertex that $\tilde b_n$ represents is given by the $n+1$ particle Feynman diagram with vertices $b_2$, $i_1 = p_1 = K_\frac{1}{2}$ inserted at each external leg, and $h = P$ used as a propagator. The corresponding $n+1$ particle Feynman diagram has the following four contributions.
\begin{equation*}
\begin{split} \
\begin{tikzpicture}[very thick,baseline= (current bounding box)]
\tikzmath{
\vx = 0;
\vy = 0;
}
\draw [photon] (\vx mm, \vy + 10 mm) node[above]{$A^\mu$} -- (\vx,\vy) node[vertex = $\gamma_\mu$] {}
;
\draw	[electron] 	(\vx -15 mm, \vy) node[left] {$\bar \psi$} -- (\vx,\vy) node[midway,below = 1.5mm] {$K_{\frac{1}{2}}$}
;
\draw	[electron]	(\vx,\vy) -- (\vx + 9 mm,\vy) node[midway,below = 1.5mm] {$P$};
\path (\vx + 9mm,\vy) -- (\vx + 18 mm,\vy) node[midway] {$\cdots$};
\draw	[electron]	(\vx + 18 mm,\vy) -- (\vx + 27 mm,\vy);
\draw [photon] (\vx + 27 mm, \vy + 10 mm) node[above]{$A^\mu$} -- (\vx + 27mm,\vy) node[vertex = $\gamma_\mu$] {}
;
\draw	[electron]	(\vx + 27 mm,\vy) -- (\vx + 36 mm,\vy) node[midway,below = 1.5mm] {$P$};
\draw [photon] (\vx + 36 mm, \vy + 10 mm) node[above]{$A^\mu$} -- (\vx + 36mm,\vy) node[vertex = $\gamma_\mu$] {}
;
\draw	[electron] 	(\vx + 36 mm, \vy)  -- (\vx + 51mm,\vy) node[midway,below = 1.5mm] {$K_{\frac{1}{2}}$} node[right] {$\psi$}
;
\end{tikzpicture} \,
\\
+ \ 
\begin{tikzpicture}[very thick,baseline= (current bounding box)]
\tikzmath{
\vx = 0;
\vy = 0;
}
\draw [dashed] (\vx mm, \vy + 10 mm) node[above]{$c$} -- (\vx,\vy) node[vertex = $i$] {}
;
\draw	[electron] 	(\vx -15 mm, \vy) node[left] {$\psi^*$} -- (\vx,\vy) node[midway,below = 1.5mm] {$K_{\frac{1}{2}}$}
;
\draw	[electron]	(\vx,\vy) -- (\vx + 9 mm,\vy) node[midway,below = 1.5mm] {$P$};
\path (\vx + 18mm,\vy) -- (\vx + 27 mm,\vy) node[midway] {$\cdots$};
\draw	[electron]	(\vx + 9 mm,\vy) -- (\vx + 18 mm,\vy);
\draw [photon] (\vx + 9 mm, \vy + 10 mm) node[above]{$A^\mu$} -- (\vx + 9mm,\vy) node[vertex = $\gamma_\mu$] {}
;
\draw	[electron]	(\vx + 27 mm,\vy) -- (\vx + 36 mm,\vy) node[midway,below = 1.5mm] {$P$};
\draw [photon] (\vx + 36 mm, \vy + 10 mm) node[above]{$A^\mu$} -- (\vx + 36mm,\vy) node[vertex = $\gamma_\mu$] {}
;
\draw	[electron] 	(\vx + 36 mm, \vy)  -- (\vx + 51mm,\vy) node[midway,below = 1.5mm] {$K_{\frac{1}{2}}$} node[right] {$\psi$}
;
\end{tikzpicture} \,
\\
+ \ 
\begin{tikzpicture}[very thick,baseline= (current bounding box)]
\tikzmath{
\vx = 0;
\vy = 0;
}
\draw [photon] (\vx mm, \vy + 10 mm) node[above]{$A^\mu$} -- (\vx,\vy) node[vertex = $\gamma_\mu$] {}
;
\draw	[electron] 	(\vx -15 mm, \vy) node[left] {$\bar \psi$} -- (\vx,\vy) node[midway,below = 1.5mm] {$K_{\frac{1}{2}}$}
;
\draw	[electron]	(\vx,\vy) -- (\vx + 9 mm,\vy) node[midway,below = 1.5mm] {$P$};
\path (\vx + 9mm,\vy) -- (\vx + 18 mm,\vy) node[midway] {$\cdots$};
\draw	[electron]	(\vx + 18 mm,\vy) -- (\vx + 27 mm,\vy);
\draw [photon] (\vx + 27 mm, \vy + 10 mm) node[above]{$A^\mu$} -- (\vx + 27mm,\vy) node[vertex = $\gamma_\mu$] {}
;
\draw	[electron]	(\vx + 27 mm,\vy) -- (\vx + 36 mm,\vy) node[midway,below = 1.5mm] {$P$};
\draw [dashed] (\vx + 36 mm, \vy + 10 mm) node[above]{$c$} -- (\vx + 36mm,\vy) node[vertex = $i$] {}
;
\draw	[electron] 	(\vx + 36 mm, \vy)  -- (\vx + 51mm,\vy) node[midway,below = 1.5mm] {$K_{\frac{1}{2}}$} node[right] {$\bar \psi^*$}
;
\end{tikzpicture}
\\
+ \ 
\begin{tikzpicture}[very thick,baseline= (current bounding box)]
\tikzmath{
\vx = 0;
\vy = 0;
}
\draw [dashed] (\vx mm, \vy + 10 mm) node[above]{$c$} -- (\vx,\vy) node[vertex = $i$] {}
;
\draw	[electron] 	(\vx -15 mm, \vy) node[left] {$\psi^*$} -- (\vx,\vy) node[midway,below = 1.5mm] {$K_{\frac{1}{2}}$}
;
\draw	[electron]	(\vx,\vy) -- (\vx + 9 mm,\vy) node[midway,below = 1.5mm] {$P$};
\path (\vx + 9mm,\vy) -- (\vx + 18 mm,\vy) node[midway] {$\cdots$};
\draw	[electron]	(\vx + 18 mm,\vy) -- (\vx + 27 mm,\vy);
\draw [photon] (\vx + 27 mm, \vy + 10 mm) node[above]{$A^\mu$} -- (\vx + 27mm,\vy) node[vertex = $\gamma_\mu$] {}
;
\draw	[electron]	(\vx + 27 mm,\vy) -- (\vx + 36 mm,\vy) node[midway,below = 1.5mm] {$P$};
\draw [dashed] (\vx + 36 mm, \vy + 10 mm) node[above]{$c$} -- (\vx + 36mm,\vy) node[vertex = $i$] {}
;
\draw	[electron] 	(\vx + 36 mm, \vy)  -- (\vx + 51mm,\vy) node[midway,below = 1.5mm] {$K_{\frac{1}{2}}$} node[right] {$\bar \psi^*$}
;
\draw (\vx,\vy - 7 mm) edge (\vx,\vy - 10 mm);
\draw (\vx,\vy - 10 mm)  edge (\vx +36mm,\vy - 10 mm);
\draw (\vx +36mm,\vy - 10 mm) edge (\vx +36mm,\vy - 7 mm);
\path (\vx,\vy - 10 mm)  -- (\vx +36mm,\vy - 10 mm) node[midway,below] {$(n-2)$ times};
\end{tikzpicture}
\end{split}
\end{equation*}
Since $h(A_\mu) = 0$, there are no internal photon lines and so the above above four diagrams are the only ones that contribute. We can simplify things further by noting that the ghost and photon interactions can be combined into vertices $V_0^\dagger$ and $V_0$, where
\begin{align} \label{V0dagger}
V_0^\dagger  =  \bar{\psi}K_{\frac{1}{2}}\cA + i\psi^* K_{\frac{1}{2}}c = & \quad \begin{tikzpicture}[very thick,baseline= (current bounding box)]
\tikzmath{
\vx = 0;
\vy = 0;
}
\draw [photon] (\vx mm, \vy + 10 mm) node[above]{$A_\mu$} -- (\vx,\vy) node[vertex = $\gamma_\mu$] {}
;
\draw	[electron] 	(\vx -15 mm, \vy) node[left] {$\bar \psi$} -- (\vx,\vy) node[midway,below = 1.5mm] {$K_{\frac{1}{2}}$}
;
\draw	[electron]	(\vx,\vy) -- (\vx + 9 mm,\vy)
;
\end{tikzpicture}
+
\begin{tikzpicture}[very thick,baseline= (current bounding box)]
\tikzmath{
\vx = 0;
\vy = 0;
}
\draw [dashed] (\vx mm, \vy + 10 mm) node[above = 2mm]{$c$} -- (\vx,\vy) node[vertex = $i$] {}
;
\draw	[electron] 	(\vx -15 mm, \vy) node[left] {$ \psi^*$} -- (\vx,\vy) node[midway,below = 1.5mm] {$K_{\frac{1}{2}}$}
;
\draw	[electron]	(\vx,\vy) -- (\vx + 9 mm,\vy)
;
\end{tikzpicture} \, ,
\\  \label{V0}
V_0 = \cA K_{\frac{1}{2}}\psi +  i c K_{\frac{1}{2}}\bar{\psi}^* = & \quad \begin{tikzpicture}[very thick,baseline= (current bounding box)]
\tikzmath{
\vx = 0;
\vy = 0;
}
\draw [photon] (\vx mm, \vy + 10 mm) node[above]{$A_\mu$} -- (\vx,\vy) node[vertex = $\gamma_\mu$] {}
;
\draw	[relectron] 	(\vx +15 mm, \vy) node[right] {$\psi$} -- (\vx,\vy) node[midway,below = 1.5mm] {$K_{\frac{1}{2}}$}
;
\draw	[relectron]	(\vx,\vy) -- (\vx - 9 mm,\vy)
;
\end{tikzpicture}
+
\begin{tikzpicture}[very thick,baseline= (current bounding box)]
\tikzmath{
\vx = 0;
\vy = 0;
}
\draw [dashed] (\vx mm, \vy + 10 mm) node[above = 2mm]{$c$} -- (\vx,\vy) node[vertex = $i$] {}
;
\draw	[relectron] 	(\vx + 15 mm, \vy) node[right] {$\bar \psi^*$} -- (\vx,\vy) node[midway,below = 1.5mm] {$K_{\frac{1}{2}}$}
;
\draw	[relectron]	(\vx,\vy) -- (\vx - 9 mm,\vy)
;
\end{tikzpicture} \, .
\end{align}
The subscript $0$ indicates that the total ghost number is zero. In a moment, we will  introduce other vertices with different ghost numbers. In addition we define $E_n = -(-\cA P)^n$. The interaction $S_{n+1}$ generating the product $b_n$ can then be written as
\begin{equation}
S_{n+1}(\tau) = \int V_0^\dagger P E_{n-3} V_0 \, ,
\end{equation}
as long as $n \ge 3$. To cover the special case $n = 2$, we also introduce
\begin{equation}\label{V1}
V_{1} = c K_\frac{1}{2}\psi \, , \quad V_{1}^\dagger = -\bar{\psi} K_\frac{1}{2} c \, ,
\end{equation}
so that 
\begin{equation}
S_3(\tau) = \int \bar{\psi} K_{\frac{1}{2}}V_0 + i \psi^* K_{\frac{1}{2}}V_{1} = \int V_0^\dagger K_{\frac{1}{2}} \psi + i V_1^\dagger K_{\frac{1}{2}} \bar{\psi}^* \, .
\end{equation}
The action $S(\tau)$ with stub length $\tau$ is then
\begin{equation}\label{stubS}
S(\tau) =  \int i\bar{\psi}\cp\psi - \frac{1}{4}F_{\mu\nu}F^{\mu\nu} - A^*_\mu \partial^\mu c + \sum_{n \ge 3} S_n(\tau) \, .
\end{equation}

\subsection{Classical Master Equation with Stubs}
As a consistency check, we will now proof that \eqref{stubS} satisfies the classical master equation. We want to stress that this does not happen by accident. The homotopy transfer theorem guarantees that the induced algebra structure is an $L_\infty$ algebra. An action generating this algebra then automatically satisfies the classical master equation. On the other hand, we only argued loosely on how such an action should look like, without a rigorous application of the homotopy transfer theorem. This makes the crosscheck necessary.

In order to check the classical master equation, we split \eqref{stubS} into three different pieces. First of all, we write
\begin{equation}
S(\tau) = S_0 + S_I(\tau) \, ,
\end{equation}
where $S_0$ contains all the terms quadratic in the fields. We further split
\begin{equation}
S_I = S_I^{(3)} + S_I^{(\geq 4)} \, ,
\end{equation}
such that $S_I^{(3)}$ contains the cubic interactions. As before, we write $Q_0 = \{S_0, -\}$ which acts as
\begin{equation}
Q_0(A_\mu) = \partial_\mu c, \quad Q_0(\bar{\psi}^*) = i\gamma^\mu \partial_\mu \psi, \quad Q_0(\psi^*) = i \partial_\mu \bar{\psi}\gamma^\mu \, .
\end{equation} 
Recall the definitions of $V_0$ and $V_0^\dagger$ given in \eqref{V0dagger} and \eqref{V0}. They satisfy 
\begin{align}
Q_0(V_0) &= \gamma^\mu \partial_\mu c K_{\frac{1}{2}}\psi + i c K_{\frac{1}{2}}\gamma^\mu \partial_\mu \psi = \gamma^\mu \partial_\mu (c K_{\frac{1}{2}}\psi) = \gamma^\mu \partial_\mu V_1 \\
Q_0(V_0^\dagger) &= - \bar{\psi}K_{\frac{1}{2}}\gamma^\mu \partial_\mu c - \partial_\mu \bar{\psi} K_{\frac{1}{2}}\gamma^\mu c = - \partial_\mu\bar{\psi}K_{\frac{1}{2}}c \gamma^\mu = - \partial_\mu V_1^\dagger \gamma^\mu \, , \nonumber
\end{align}
where $V_1$ and $V_1^\dagger$ are defined in \eqref{V1}. We further have
\begin{align}\label{cv0}
c V_0 &= \slashed{A}V_1, \quad V_0^\dagger c = V_1^\dagger \slashed{A}.
\end{align}
Acting on $S_I$ with $Q_0$ we find
\begin{subequations}\label{e3}
\begin{align}
Q_0(S_I) =& -Q_0  (V^\dagger_0 P \sum_{n \ge 0} (-\slashed{A} P)^n V_0 )\nonumber\\ =&
  (\partial_\mu V_1^\dagger) \gamma^\mu  P \sum_{n \ge 0} (-\slashed{A} P)^n V_0 + V_0^\dagger \sum_{n \ge 0} (-\slashed{A} P)^n \slashed{\partial} V_1\label{e3a} \\
&-V^\dagger_0  \sum_{n \ge 0} (-P\slashed{A})^n P (\slashed{\partial} c) P \sum_{m \ge 0} (-\slashed{A} P)^m V_0 \,  .\label{e3b}
\end{align}
\end{subequations}
In the first step we used that $Q_0(S_I^{(3)}) = 0$. For convenience, we no longer explicitly display the space time integration. To continue we use the identity
\begin{equation}\label{iddc}
P  (\slashed{\partial} c) P = -i(1 -K) c P + i P c (1 -K) \, .
\end{equation}
We first focus on part in (\ref{e3b}) resulting from the identity operator $1\!\!1$ in (\ref{iddc}). This contributes a term 
\begin{equation}\label{e5}
-i V_1^\dagger \sum_{n \ge 1} (-\slashed{A} P)^n V_0 + i V_0^\dagger \sum_{n \ge 1} (-\slashed{A} P)^n V_1 \, ,
\end{equation}
where we have used (\ref{cv0}). 
There is a similar contribution from (\ref{e3a}) as can be seen using 
\begin{equation}\label{dv1}
\partial_\mu V_1^\dagger \gamma^\mu P = i V_1^\dagger(1 -K), \quad P \slashed {\partial} V_1 = -i (1 - K) V_1 \, .
\end{equation}
and focusing on the pieces in (\ref{e3a}) originating from the identity operator $1\!\!1$, 
\begin{equation}
i V_1^\dagger \sum_{n \ge 0} (\slashed A P)^n V_0 - i V_0^\dagger \sum_{n \ge 0} (P \slashed A)^n V_1 \, ,
\end{equation}
which adds up with \eqref{e5} to
\begin{equation}
i(V_1^\dagger V_0 - V_0^\dagger V_1) = 0 \, .
\end{equation}
In summary, the only terms left are those which involve $K$ upon using (\ref{cv0}) and (\ref{dv1}), that is, 
\begin{subequations}\label{e6}
\begin{align}
Q_0(S_I))=&-i V_0^\dagger \sum_{n \ge 0} (-P \slashed A)^n K c P \sum_{m \ge 0}(-\slashed A P)^m V_0 + i V_0^\dagger \sum_{n \ge 0} (-P \slashed A)^n P c K \sum_{m \ge 0}(-\slashed A P)^m V_0 \, \label{e6a}\\
&- i V_1^\dagger K \sum_{n \ge 0} (-\slashed A P)^n V_0 + i V_0^\dagger \sum_{n \ge 0} (-P \slashed A)^n K V_1 \, .\label{e6b}
\end{align}
\end{subequations}

We now turn to the anti-bracket of $S_I$ with itself. For this it is useful to note that
\begin{equation}
\frac{1}{2}\{S_I,S_I\} = \frac{\partial_r S_I}{\partial \psi^*}\frac{\partial_l S_I}{\partial \psi} + \frac{\partial_r S_I}{\partial \bar{\psi}^*}\frac{\partial_l S_I}{\partial \bar{\psi}} = - \frac{\partial_r S_I}{\partial \psi}\frac{\partial_l S_I}{\partial \psi^*} + \frac{\partial_r S_I}{\partial \bar{\psi}^*}\frac{\partial_l S_I}{\partial \bar{\psi}} \, .
\end{equation} 
The later form is convenient as, since $S_I$ always starts with either $\psi^*$ or $\bar{\psi}$ and ends with $\psi$ or $\bar{\psi}^*$, all functional derivatives only act on the first object they encounter. This gives
\begin{align}
\frac{1}{2}\{S_I^{(\ge 4)},S_I^{(\ge 4)}\} =& +i V_0^\dagger \sum_{n \ge 1} (-P\slashed A)^n K c P \sum_{m \ge 0} (-\slashed A P)^m V_0 \nonumber\\
&- i V_0^\dagger \sum_{n \ge 0} (-P\slashed A)^n PcK \sum_{m \ge 1} (-\slashed A P)^m V_0 \, . 
\end{align}
Adding this to \eqref{e6a} we are left with
\begin{equation}\label{eq8}
-i V_0^\dagger KcP \sum_{n \ge 0} (-\slashed A P)^n V_0 + i V_0^\dagger \sum_{n \ge 0} (-P\slashed A)^n PcK V_0 \, .
\end{equation}
Finally, adding to this
\begin{align}
\{S_I^{(\ge 4)},S_I^{(3)}\} =& i V_0^\dagger KcP \sum_{n \ge 0} (-\slashed A P)^n V_0 + i V_1^\dagger K \sum_{n \ge 1} (-\slashed A P)^n V_0 \nonumber\\
							&- i V_0^\dagger \sum_{n \ge 0} (-P \slashed A)^n PcK V_0 - i V_0^\dagger  \sum_{n \ge 1}(-P\slashed A)^n K V_1 \, 
\end{align}
cancels \eqref{eq8} as well as the terms in \eqref{e6b} with $n\neq 0$, so that we are left with the $n=0$ contribution, 
\begin{equation}
-i V_1^\dagger K V_0 + i V_0^\dagger K V_1 \, 
\end{equation}
from \eqref{e6b}, which is, in turn,  cancelled by 
\begin{equation}
\{S_I^{(3)},S_I^{(3)}\} = i V_1^\dagger K V_0 - i V_0^\dagger K V_1 \, .
\end{equation}
Thus, we have shown that
\begin{equation}\label{pbv}
Q_0(S_I) + \frac{1}{2}\{S_I,S_I\} = 0 \, 
\end{equation}
by explicit calculation. In the next section we will give a more concise derivation of this by construction of a generating function so that $S_I$ can be written as a canonical transformation. 

Before moving on, let us comment on possible generalisations of this result. The extension to a massive Dirac field is straight forward. One simply has to replace $\square$ by $\square+m^2$ and $\slashed\partial$ by $\slashed\partial+i\,m$ in the definition of $K$ and $P$. Then \eqref{stubS} gives the correct BV-regularization to all orders, for masssive $QED$ in $n$ dimensions. The non-abelian generalization is more subtle. First of all, the  BV-action \eqref{SS1} receives an extra contribution to the Yang-Mills action, that is 
\begin{align}\label{bana}
    S_1=i\bar{\psi} \bc \bar{\psi}^* + i \psi^* \bc \psi  - A^{r*}_\mu (\partial^\mu c^r-if^r_{st}A^{s\mu}c^t) \,.
\end{align}
where $\bc=c^r\tau^r$, $\tau^r\in \mathfrak{g}$ and $f^r_{st}$ are the structure constants of the Lie algebra.  In the derivation of \eqref{pbv} we have used that the ghost $c$ commutes with the vector potential $A_\mu$. In the non-abelian version this produces some extra terms involving $[c,A_\mu]$. However, these are cancelled by the new terms in the master equation arising from the last term in \eqref{bana}. As a result, \eqref{stubS} is again the correct regularization of the fermionic action in the non-abelian case. However, to last term in \eqref{bana} the full BV-action requires further regularization in order for the BV-operator $\Delta$ to be well-defined. This amounts to adding stubs to $A_\mu$ as well. While these stubs can be added by the same linear transformation implemented by the action of $K_\frac{1}{2}$, the additional terms $S^{(n>3)}$ will be more complicated and we do not have a closed form for them.

\section{Stubs as a Canonical Transformation}\label{GF}
In \eqref{GenFun} we described an efficient way on how to parametrize deformations of a given action subject to the quantum BV-equation via a canonical transformation. Given a generating function $R(\tau)$, finding the deformed action $R(\tau)$ amounts to solving the equation
\begin{align}\label{ct1}
    \partial_\tau S(\tau)=\{S(\tau),R(\tau)\}-i\hbar\Delta R(\tau) \, .
\end{align}
Below we want to find an $R(\tau)$ that generates the action \eqref{stubS} at the classical level, i.e. when $\hbar = 0$. In this case, we can think of $X = \{R(\tau),-\}$ as a vector field generating the field redefinitions that allow us to pass between actions $S(\tau)$ of arbitrary stub parameter $\tau$. An important observation is that in our case the quadratic terms in $S(\tau)$ are independent of $\tau$. The field redefinition should therefore be of the form
\begin{equation}\label{quadraticdeformation}
\psi \mapsto \psi + \mathcal{O}(\psi^2) \, .
\end{equation}
This in turn implies that the generating function $R(\tau)$ should start at cubic order. 

The map \eqref{quadraticdeformation} is different from the field redefinition determined by the homotopy transfer theorem, where the map $\psi \mapsto  i(\psi)$ contains a linear term  $i_1(\psi) = K_{\frac{1}{2}}\psi$. This does not lead to a contradiction however, since $i$ is a map between equations of motion, while the map in \eqref{quadraticdeformation} is a map between actions.  The latter makes it also clear that the stub regularization described here is different from the cut-off propagator used in exact renormalization group flow as in \cite{Igarashi:2009tj}, for example. 

\subsection{The Generating Functional to Lowest Order}\label{CT}
We begin by constructing $R(\tau)$ to lowest (cubic) order. To keep the formulas concise, we introduce two more symbols,
\begin{equation}
V_{-1} := i K_{\frac{1}{2}} \slashed \partial  \bar{\psi}^* \quad\text{and}\quad V^\dagger_{-1} : = i \slashed \partial \psi^* K_{\frac{1}{2}} \, ,
\end{equation}
which satisfy
\begin{equation}
Q_0(V_{-1}) = - K_{\frac{1}{2}} \square  \psi \quad\text{and}\quad Q_0(V_{-1}^\dagger) = - \square \bar{\psi} K_{\frac{1}{2}} \, 
\end{equation}
respectively. To lowest order in the deformation a generating function that reproduces $S_I^{(3)}$ in \eqref{stubS}, upon integration of \eqref{ct1}, is then given by
\begin{align}\label{r3}
R^{(3)}(\tau) 	=& -\frac{1}{2}V^\dagger_0 V_{-1} + \frac{1}{2}V^\dagger_{-1} V_0 \, \nonumber\\
				=& - i\frac{1}{2}\bar{\psi}K_{\frac{1}{2}}\slashed A K_{\frac{1}{2}} \slashed \partial \bar{\psi}^* + \frac{1}{2} \psi^* K_{\frac{1}{2}} c K_{\frac{1}{2}} \slashed \partial \bar{\psi}^* \nonumber\\
				&+\frac{1}{2}i \partial_\mu \psi^* \gamma^\mu  K_{\frac{1}{2}} \slashed A K_{\frac{1}{2}} \psi - \frac{1}{2} \partial_\mu \psi^* \gamma^\mu K_{\frac{1}{2}} c K_{\frac{1}{2}}\bar{\psi}^*
\end{align}
Indeed, using $\partial_\tau K_{\frac{1}{2}} = - \frac{1}{2} \square K_{\frac{1}{2}}$ we get from \eqref{stubS}
\begin{equation}
\partial_\tau S_I^{(3)} = -\frac{1}{2} \square \bar{\psi} K_{\frac{1}{2}} V_0 - i\frac{1}{2} \square \psi^* K_{\frac{1}{2}} V_1 - \frac{1}{2}V_0^\dagger K_{\frac{1}{2}}\square \psi - i \frac{1}{2}V_1^\dagger K_{\frac{1}{2}} \square \bar{\psi}^* \, . 
\end{equation}
On the other hand,
\begin{align}
Q_0 R^{(3)}(\tau) &= i\frac{1}{2} \slashed \partial V_1^\dagger K_{\frac{1}{2}} \slashed \partial \bar{\psi}^* - \frac{1}{2} V_0^\dagger K_{\frac{1}{2}} \square \psi - \frac{1}{2}\square \bar{\psi} K_{\frac{1}{2}} V_0 + \frac{1}{2} i \slashed \partial \psi^* K_{\frac{1}{2}} \slashed \partial V_1 \\
&= -\frac{1}{2} i V_1^\dagger K_{\frac{1}{2}} \square \bar{\psi}^* - \frac{1}{2} V_0^\dagger K_{\frac{1}{2}} \square \psi - \frac{1}{2}\square \bar{\psi} K_{\frac{1}{2}} V_0 - \frac{1}{2} i \square \psi^* K_{\frac{1}{2}} V_1 \, .
\end{align}
Hence,
\begin{equation}\label{ps31}
\partial_\tau S_I^{(3)} = Q_0 R^{(3)}(\tau) \, .
\end{equation}

\subsection{The Generating Function to all Orders}

Starting from \eqref{r3} we may make an educated  guess the generating function to all orders. We claim that higher order terms are obtained by a repeated insertions of $-P\cA$, similar to the way higher order vertices in $S(\tau)$ are generated. To simplify the formulas we define $E = \sum_{n \ge 0} (-\cA P)^n$ and $E^\dagger = \sum_{n \ge 0} (-P \cA)^n$. Then our Ansatz reads
\begin{equation}\label{GR}
R(\tau) = - \frac{1}{2}V_0^\dagger E^\dagger V_{-1} + \frac{1}{2} V_{-1}^\dagger E V_0 \, .
\end{equation}

In order to proof that $R(\tau)$ generates $S(\tau)$, we first provide two useful intermediate steps which are straightforward to check. We have
\begin{equation}\label{ident1}
\{EV_0,S\} = -\cp(V_1 - cPE V_0)\, , \quad \{S,V_0^\dagger E^\dagger\} = (V_1^\dagger - V_0^\dagger E^\dagger P c)\cp \, .
\end{equation}
We also compute
\begin{equation}\label{ident2}
 -\frac{1}{2}V_0^\dagger E^\dagger\{V_{-1},S\} = - V_0^\dagger E^\dagger \partial_\tau K_\frac{1}{2} \psi - \frac{1}{2} V_0^\dagger \partial_\tau(P E) V_0 \, ,
\end{equation}
and similarly
\begin{equation}\label{ident3}
\frac{1}{2} \{S,V_{-1}^\dagger\}E V_0 = \bar{\psi}\partial_\tau K_\frac{1}{2}E V_0 - \frac{1}{2} V_0^\dagger \partial_\tau(P E) V_0 \, .
\end{equation}
Let is now consider equation \eqref{ct1} defining canonical transformations. For $\hbar = 0$, we have the right hand side,
\begin{equation}
\{S,R\} = - \frac{1}{2}\{S,V_0^\dagger E_0^\dagger V_{-1}\} + \frac{1}{2}\{S,V_{-1}^\dagger EV_0\} \, .
\end{equation}
Using the identities (\ref{ident1}-\ref{ident3}), one can then show that
\begin{equation}
\{S,\frac{1}{2} V_{-1}^\dagger E V_0\} =  \partial_\tau(\bar{\psi}K_{\frac{1}{2}})V_0 + i \partial_\tau(\psi^*K_{\frac{1}{2}})V_1 - (\partial_\tau V_0)PE V_0 - \frac{1}{2} V_0^\dagger \partial_\tau (PE)V_0
\end{equation}
and
\begin{equation}
- \{S,\frac{1}{2}V_0 E^\dagger V_{-1}\} =  V_0^\dagger \partial_\tau( K_{\frac{1}{2}} \psi) + i V_1^\dagger  \partial_\tau (K_{\frac{1}{2}} \bar{\psi}^*) - V_0^\dagger E^\dagger P \partial_\tau V_0 - \frac{1}{2} V_0^\dagger \partial_\tau (PE)V_0 \, .
\end{equation}
This combines to
\begin{equation}
\{S,R\} = \partial_\tau S \, .
\end{equation}

To conclude we note that the existence of a generating function provides another proof of the fact that $S(\tau)$ satisfies the classical master equation for all $\tau$, since we already know that it does so for $\tau = 0$. At zero stub length, we just recover the original cubic BV action of QED.

\subsection{Quantum Correction Induced by the Generating Functional}

So far we considered the classical flow $\partial_\tau S =\{S,R\}$ induced by the generating functional $R(\tau)$. At the quantum level, there is an additional contribution due to the change of the path integral measure. They enter in the flow equation through,
\begin{equation}\label{qb}
\partial_\tau S = \{S,R\} - i \hbar \Delta R \, .
\end{equation}
The terms in $R$ that give a non-zero contribution when acting with $\Delta$ are
\begin{equation}
-\frac{i}{2}\bar{\psi} K_{\frac{1}{2}} \cA E^\dagger K_{\frac{1}{2}}\cp \bar{\psi}^* - \frac{i}{2}\psi^* K_{\frac{1}{2}} \cp E \cA K_{\frac{1}{2}} \psi \, .
\end{equation}
Acting with the Laplacian we find
\begin{align}\label{PIQ}
-\frac{i}{2}\tr(\cA E^\dagger \cp K) - \frac{i}{2}\tr(\cp K  E \cA) &= -i \tr(E \cA \cp K) = \tr (E \cA \dot P) \\
&= -\partial_\tau \tr \sum_{n \ge 1}\frac{(-)^n}{n} (P \cA)^n  \,,\nonumber
\end{align}
where the trace is over spinor indices.  Therefore, the quantum correction to $S(\tau)$ generated by $R(\tau)$ reads
\begin{equation}\label{Iq}
I_q := i \hbar \tr \sum_{n \ge 1} \frac{(-)^n}{n}(P \cA)^n \, .
\end{equation}
Note that the tadpole at $n = 1$ vanishes since it is a total derivative. On the other hand, $I_q$ is ultraviolet divergent. This can be seen by noting that for for $\tau\to\infty$, $I_q$ reproduces the sum over one-loop diagrams with $n$ external photons. Thus, in order to define $I_q$, we have to make a suitable subtraction. If we denote the renormalized $I_q$ by $I^{ren}_q$ and assume that the counter term is $Q_0$-invariant one  
then shows that $I^{ren}_q$ is indeed the addition required for $S(\tau)$ to satisfy the full quantum master equation. Indeed, on one hand we have
\begin{equation}
i\Delta S = i\Delta S_I =  \tr(K \cA P E c) -  \tr (K c P E \cA) \, .
\end{equation}
while on the other hand,
\begin{align}\label{IS}
\frac1\hbar Q_0(I^{ren}_q)=
\frac1\hbar Q_0(I_q) &= -i \tr (P \cp c P \cA E^\dagger) =  \tr((1-K)c P \cA E^\dagger) - \tr (P c (1-K) \cA E^\dagger)\nonumber \\
&= -\tr (K c P E \cA) + \tr (P c K  E \cA) + \tr (c E^\dagger P \cA) - \tr(Pc \cA E^\dagger) \nonumber\\
&= i\Delta S \, .
\end{align}
 where we  assumed cyclicity of the trace to conclude that the action $S_q = S + I_q$ satisfies the quantum master equation and is therefore non-anomalous. 

Given the exponential suppression, at large (Euclidean) momentum, terms containing $K$ are UV-convergent so that the desired cyclicity holds. Recalling from \eqref{PA}
\begin{align}\label{PAE}
    P = -i \cp \int_0^\tau \text d s \,  e^{-s \square}= \frac{i\cp}{\square}(e^{-\tau \square}-1)\,,
\end{align} 
we then see that terms containing only $P$'s but no $K$'s may need extra attention. In view of the axial theory, involving $\gamma_5$, described in the next section we use Pauli-Villars regularization, that is, we add to \eqref{Iq} contributions from fermions with masses $M_i$ suitable statistics and propagator 
\begin{align}\label{PAEM}
    P^{M_i} = \frac{i\cp-M_i}{\square+M_i^2}(e^{-\tau (\square+M_i^2)}-1)\stackrel{M_i\to\infty}{\simeq}-\frac{i\cp-M_i}{\square+M_i^2}\,.
\end{align} 
Thus the subtraction is identical to that for the one-loop diagrams with $n$ external photons in QED without stubs. For the PV-regulated $I_q$ the cyclic property assumed in \eqref{IS} then holds. So we are left with the action action of $Q_0$ on the contribution of the PV-fermions which amounts to testing the classical gauge invariance of this amplitude. We can then use the known gauge invariance of the PV-regulator to conclude that $Q_0$ vanishes on these extra contributions. 


In closing this section we note that the fact that quantum correction come from a generating function, this does not by itself imply that $S_q$ satisfies the quantum master equation. Only if $S_q(\tau)$ satisfies the quantum master equation at one particular value, $\tau = \tau $, the existence of the generating function guarantees that the master equation is satisfied for \emph{all} $\tau$. In contrast to the classical case, we cannot just deduce this from $S_q(\tau = 0)$. Indeed since $S(\tau = 0) = 0$, we reproduce the cubic QED action $S $ \eqref{QEDaction}(on which $\Delta$ is singular) in that limit. Furthermore, $Q_0(I_q) = i \hbar\Delta(S_I)$ is non-zero in the limit $\tau \rightarrow 0$. Thus, there still appears a quantum correction to the classical gauge transformations generated by $S $, even when we send the stub length to zero. This correction is not visible in $S(\tau =  0)=S_{cl}$.

\section{Axial Theory}\label{aqed}

In this section we state the results derived in the previous section for QED with an axial gauge symmetry, without repeating all the computations. The analog of action \eqref{stubS} for the axial theory with stub length $\tau$ is given by $S_5(\tau)=S_0+S_{I5}$ with
\begin{align}
S_{I5}(\tau) &= -V_0^\dagger \gamma_5 \psi + i V_1^\dagger \gamma_5 \bar{\psi}^* - V_0^\dagger PE_5 V_0  
  = \bar{\psi}K_{\frac{1}{2}}\gamma_5 V_0 - i \psi^* K_{\frac{1}{2}} \gamma_5 V_1  - V_0^\dagger E^\dagger_5 P V_0 \, .
\end{align}
where
\begin{equation}
E_5 = \sum_{n \ge 0} (-\gamma_5 \cA P)^n \, , \quad E_5^\dagger = \sum_{n \ge 0} (-P \gamma_5 \cA)^n \, .
\end{equation}
The modification of of the generating function $R_5(\tau)$ that generates $S_5(\tau)$ at the classical level is again easily deduced with  
\begin{equation}
R_5(\tau) = \frac{1}{2} V_{-1}^\dagger \gamma_5 E_5 V_0 + \frac{1}{2}V_0^\dagger E_5^\dagger \gamma_5 V_{-1} \, .
\end{equation}
In order to obtain the quantum corrections to the BV action, as before, using \eqref{qb} we evaluate the Laplacian on $R_5(\tau)$, 
\begin{equation}
i\hbar\Delta(R_5) = i \hbar \tr(\partial_\tau P E_5 \gamma_5 \cA)\, .
\end{equation}
This integrates to
\begin{equation}\label{IQ5}
I_{5,q}(\tau) =  i\hbar \tr\sum_{n \ge 1} \frac{(-1)^n}{n}(\gamma_5 \cA P)^n \equiv \sum_{n \ge 1} I_n\, .
\end{equation}
To continue, in order to guarantee UV-convergence of this expression and thereby cyclicity of the trace we may again add contributions Pauli-Villars fermions with masses $M_i$ to \eqref{IQ5}. 
To see whether $S_5 + I_{5,q}$ regularized in this way solves the quantum master equation \eqref{QME} we note that the massles fermions constribute to the BV-bracket as 
\begin{align}\label{QQC}
\frac{1}{2}\{S_5 + I_{5,q},S_5 +I_{5,q}\}   &= - i\hbar \tr (\gamma_5 \cp c P E_5)= i \hbar \tr (\cA P \cp c P E_5) \nonumber \\
                            &= \hbar \tr (\cA (1-K) c P E_5) -  \hbar \tr (\cA P c (1-K) E_5) \,,
\end{align}
which then indeed cancels against 
\begin{equation}
i\hbar \Delta (S_5 + I_{5,q}) = \hbar\tr (KcPE_5\cA) - \hbar\tr (\cA P E_5 c K) \, .
\end{equation}
The price we pay for cyclicity, used above, are the extra contributions to \eqref{QQC} from the Pauli-Villars fermions. At linear order in $\cA$, in momentum space, there is a contribution 
\begin{align}
    - \hbar \int \frac{d^4k}{(2\pi)^4}\tr_D &\left(\gamma_5\tilde c(p)\dP\frac{\dk-M_{i}}{k^2-M_{i}^2}\gamma_5 \dA(-p)\frac{\dP+\dk-M_{i}}{(p+k)^2-M_{i}^2}\right)\nonumber\\
    &=- \hbar \tilde c(p)\int \frac{d^4k}{(2\pi)^4}\tr_D \left( \tilde\dA(-p)\frac{\dP+\dk-M_{i}}{(p+k)^2-M_{i}^2}-  \tilde\dA(-p)\frac{\dk-M_{i}}{k^2-M_{i}^2}\right.\nonumber\\
    &\qquad\qquad\qquad\qquad\qquad\qquad\qquad\left.+2M_i \frac{\dk+M_{i}}{k^2-M_{i}^2}\tilde\dA(-p)\frac{\dP+\dk-M_{i}}{(p+k)^2-M_{i}^2}\right)\nonumber\\
    &= 4M_i^2\hbar \tilde c(p)p\cdot\tilde A(-p)\int \frac{d^4k}{(2\pi)^4}\,\frac{1}{(k^2-M_{i}^2)[(p+k)^2-M_{i}^2]}\,,
\end{align}
which, in turn, can be shown to vanish (e.g. using a parametric representation as in  \cite{Itzykson:1980rh}). 

At quadratic order in $\cA$, the massive fermions give a contribution 
\begin{align}
    &=- \hbar \int \frac{d^4k}{(2\pi)^4}\tr_D \left(\gamma_5\tilde c(p)\dP\frac{\dk-M_{i}}{k^2-M_{i}^2} \dA(r)\frac{\dk-\slashed{r}+M_{i}}{(k-{r})^2-M_{i}^2}\dA(-r-p)\frac{\dP+\dk-M_{i}}{(p+k)^2-M_{i}^2}\right)\,,\nonumber
    \end{align}
familiar from the calculation of the triangle diagram in QED for the axial anomaly with the difference that the sign in front of $M_i$ is alternated as a consequence of the two extra $\gamma_5$ for the external photons. We recall result for the triangle amplitude  (e.g. \cite{Bilal:2008qx})
\begin{align}
    8i\hbar M_i^2\int \frac{d^4k}{(2\pi)^4} \frac{\tilde c(p)\epsilon^{\mu\nu\lambda\rho}A_\mu(r) r_\nu A_\lambda(-r-p) p_\rho}{[k^2-M_{i}^2][(k-{r})^2-M_{i}^2][(p+k)^2-M_{i}^2]}
\end{align}
which for axial QED is then modified to 
\begin{align}
    8i\hbar M_i^2\int &\frac{d^4k}{(2\pi)^4} \frac{\tilde c(p)\epsilon^{\mu\nu\lambda\rho}A_\mu(r) (r_\nu-2k_\nu) A_\lambda(-r-p) p_\rho}{[k^2-M_{i}^2][(k-{r})^2-M_{i}^2][(p+k)^2-M_{i}^2]}\\
    &\qquad\qquad\qquad\qquad\qquad\qquad\qquad\qquad=\frac{i\hbar}{12}\tilde c(p)\epsilon^{\mu\nu\lambda\rho}A_\mu(r) r_\nu A_\lambda(-r-p) p_\rho\,.\nonumber
\end{align}
Thus 
\begin{align}\label{IS2}
 Q_0(I^{ren}_q)=\frac{i\pi\hbar}{6}\int  c\;\epsilon^{\mu\nu\lambda\rho}F_{\mu\nu}F_{\lambda\rho}\;d^4 x\,,
\end{align}
making the anomaly of axial QED manifest. It is satisfying to observe that the actual calculation of the anomaly can be reduced to the standard triangle diagram amplitude in the undeformed theory. 


\section{Conclusion}
In this note we revisited the regularization of the BV-Laplacian and the calculation of anomalies. We saw that this can be done at the price of passing through a non-polynomial action akin of what we are used to from string field theory. In the simple example presented here this action can be found explicitly to all orders in the deformation. However, to analyze possible anomalies the knowledge of the the complete non-polynomial action is not necessary. Rather anomalies are related to the  non-invariance, under classical gauge transformations, of a small number of quasi-local quantum effective vertices, which reduces to anomalous Ward-identities for the theory in question. The latter can be analyzed without reference to the regularization of the BV-Laplacian $\Delta$. Disentangling these two aspects and the complete non-local, quantum corrected BV-action for QED to all orders, together with the corresponding generating functional $R$, are the main results of this note. 

Concerning the first point one might argue that since the quantum effective vertices $I_q$ are obtained by acting with $\Delta$ on $R$, regularization of the former again mixes the definition of $\Delta$ with the (non-universal) regularization procedure through $I_q\sim(\Delta R)_{reg}$. However, this not so since, as is clear from \eqref{PIQ}, $\Delta R$ produces the $\tau$ derivative of $I_q$ which is UV-finite. Thus regularization enters only through the integration constant of the flow equation \eqref{qb}. This is made manifest in our approach. 

As for the second result, that is, having a simple generating function for the complete quantum BV-action for QED one might wonder if this could be of use elsewhere, in particular in string theory with the fermions replaced by the Ramond sector open super string field theory. The string extension of $R$ in \eqref{GR} is not hard to define. Graphically it is of the form
\begin{figure}[h!]
\centering
  \includegraphics[width=.6\textwidth]{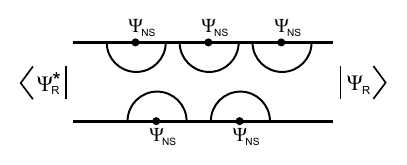}
   \caption{A possible example for the analog of the generating function in open super string field theory. We represent it as a strip with string fields $\Psi_\text{R}$ in the Ramond sector as in and out states. Fields in the Neveu Schwarz sector are inserted using vertex operators $\Psi_{\text{NS}}$.}
\end{figure}\\
Perhaps there is some use of it when adding stubs to open super string field theory. 

In closing let us mention another approach to regularization of the quantum vertices $I_q$ and $I_{q,5}$ which is more in the spirit of string field theory. Instead of integrating the stub parameter in $P$ all the way to zero, we could cut it off at some small number $\varepsilon$. The quantum master equation is then satisfied up to a term proportional to
\begin{equation}\label{epsiloncutoff}
    \tr (K(\varepsilon) cP \cA E^\dagger) - \tr (Pc K(\varepsilon) \cA E^\dagger) \, .
\end{equation}
The quantum master equation can then be satisfied if there is a correction $I_q \mapsto I_q + \delta I_q$ such that $Q_0(\delta I_q)$ cancels the above contribution. In string field theory, the cutoff $\varepsilon$ determines where on the moduli space we interpret open string loops in terms of closed strings. The presence of the closed string then takes care of the contribution similar to the one in \eqref{epsiloncutoff} via the Green-Schwarz mechanism. 

 \section*{Acknowledgments:}
We would like to thank G. Barnich for a motivating discussions during  the Solvay Workshop on 'Higher Spin Gauge Theories, Topological Field Theory and Deformation Quantization’ and T. Morris for comments on the draft and canonical transformations during the Kyoto workshop on 'Homotopy Algebra of Quantum Field Theory and Its Application'. This work was funded by the Excellence Cluster Origins of the DFG under Germany’s Excellence Strategy EXC-2094 390783311 and the European Research Council (ERC) under the 
European Union’s Horizon 2020 research and innovation programme (grant agreement No 771862).

\bibliographystyle{unsrt}
\bibliography{ref.bib}

\end{document}